\documentclass[aps,pra,twocolumn,bibnotes]{revtex4-1}
\usepackage{graphicx} 
\usepackage[caption=false]{subfig} 
\begin{document}
\title{Fully tunable and switchable coupler for photonic routing in quantum detection and modulation}

\author{Vojt\v{e}ch \v{S}varc}
\affiliation{Department of Optics, Palack\'{y} University, 17. listopadu 12, 77146 Olomouc, Czech Republic}

\author{Martina Nov\'{a}kov\'{a}}
\affiliation{Department of Optics, Palack\'{y} University, 17. listopadu 12, 77146 Olomouc, Czech Republic}

\author{Glib Mazin}
\affiliation{Department of Optics, Palack\'{y} University, 17. listopadu 12, 77146 Olomouc, Czech Republic}

\author{Miroslav Je\v{z}ek}
\email{jezek@optics.upol.cz}
\affiliation{Department of Optics, Palack\'{y} University, 17. listopadu 12, 77146 Olomouc, Czech Republic} 

\begin{abstract}
Photonic routing is a key building block of many optical applications challenging its development. We report a 2$\times$2 photonic coupler with splitting ratio switchable by a low-voltage electronic signal with 10~GHz bandwidth and tens of nanoseconds latency.
The coupler can operate at any splitting ratio ranging from 0:100 to 100:0 with the extinction ratio of 26 dB in optical bandwidth of 1.3 THz. We show sub-nanosecond switching between arbitrary coupling regimes including balanced 50:50 beam splitter, 0:100 switch, and a photonic tap.
The core of the device is based on Mach-Zehnder interferometer in a dual-wavelength configuration allowing real-time phase lock with long-term sub-degree stability at single-photon level.
Using the reported coupler, we demonstrate for the first time the perfectly balanced time-multiplexed device for photon-number-resolving detectors and also the active preparation of a photonic temporal qudit state up to four time bins.
Verified long-term stable operation of the coupler at the single photon level makes it suitable for wide application range in quantum information processing and quantum optics in general.
\end{abstract}

\maketitle

Fast splitting, switching, and routing of light are critical tools of photonic technology in the rapidly developing fields of optical communication and optical information processing including demanding applications like quantum cryptography \cite{Gauthier2018}, neuromorphic computing \cite{Englund2017,Bhaskaran2017}, photonic simulations \cite{Silberhorn2012}, scalable boson sampling \cite{Rohde2014,Pan2017full}, universal quantum computing \cite{Furusawa2017}, entanglement synthesizing \cite{takeda2019demand}, and photon counting \cite{Silberhorn2019}. In the last few years, high-efficiency single-photon generation has been demonstrated employing active time multiplexing \cite{Eggleton2016,OBrien2016,Kwiat2018}. Optical switching has also facilitated a recent pioneering demonstration of postselection-loophole-free violation of Bell's inequality with genuine time-bin entanglement \cite{Vallone2018}.

The most advanced modulation, processing, and detection schemes require ultra-low latency between the control signal and the switch response, together with a large bandwidth and high extinction. Furthermore, the continuous tunability with arbitrary splitting ratio is required
\cite{Furusawa2017,Gauthier2018}. The universal routing device would also be able to coherently superpose two incident signals acting as a coupler {\em switchable} between various splitting ratios. Free-space polarization-based switchable couplers using Pockels cells \cite{Pan2017full,Silberhorn2019,takeda2019demand} feature ultra-low loss, however, they require high switching voltage. Consequently, they cannot provide the continuous tunability and it is very demanding to reach ultra-low latency.
An alternative technique utilizes cross-phase modulation in Sagnac interferometer driven by an auxiliary strong optical pulse \cite{Kumar2011}. This approach is polarization-insensitive, exhibits medium loss and high speed, however, the latency is high, making its usage impractical in most loop-based schemes and temporal multiplexing in general. Another approach employs electro-optic phase modulators (EOMs) in Mach-Zehnder interferometer (MZI) \cite{Zeilinger2011,Vallone2018}. Using integrated EOMs instead of free-space modulators we can reach ultra-high speed and low latency at the cost of increased loss.

The MZI operates as a $2\times2$ variable beam splitter and allows the continuous tunning of its splitting ratio. The main drawbacks of MZI based switchable coupler are the extinction ratio limited by visibility of the MZI and its phase instability causing the drift of the splitting ratio. The visibility optimization is particularly challenging in the case of a spectrally broad signal, like short optical pulses and the majority of single-photon sources, and with dispersion elements utilized as a part of the MZI. The phase stability issue can be addressed by active phase locking, though it is notoriously uneasy at a single-photon level or with fluctuating input signal.

In this Letter, we present a low-latency switchable coupler employing a high-visibility fiber MZI. An auxiliary light beam is injected to the MZI, co-propagates with a single-photon signal, and enables real-time continuous phase locking with a unique sub-1 deg long-term stability. The picowatt-level auxiliary beam is wavelength separated with virtually no crosstalk to the signal.
We demonstrate fast switching of the coupler by changing its operation between any splitting ratios in a fraction of nanosecond. The splitting ratio is controlled using low-voltage electronic signal compatible with the output of the majority of photodetectors, which is crucial for utilization of the coupler in optical feedback and feedforward circuits. We show outstanding performance of the reported device in two demanding applications, namely a balanced time-multiplexed device for photon-number-resolving detectors and an active preparation of a photonic time-bin encoded 4-level state with time-bin separation in the range of tens of nanoseconds.

\begin{figure}[t]
\centering
\includegraphics[width=\linewidth]{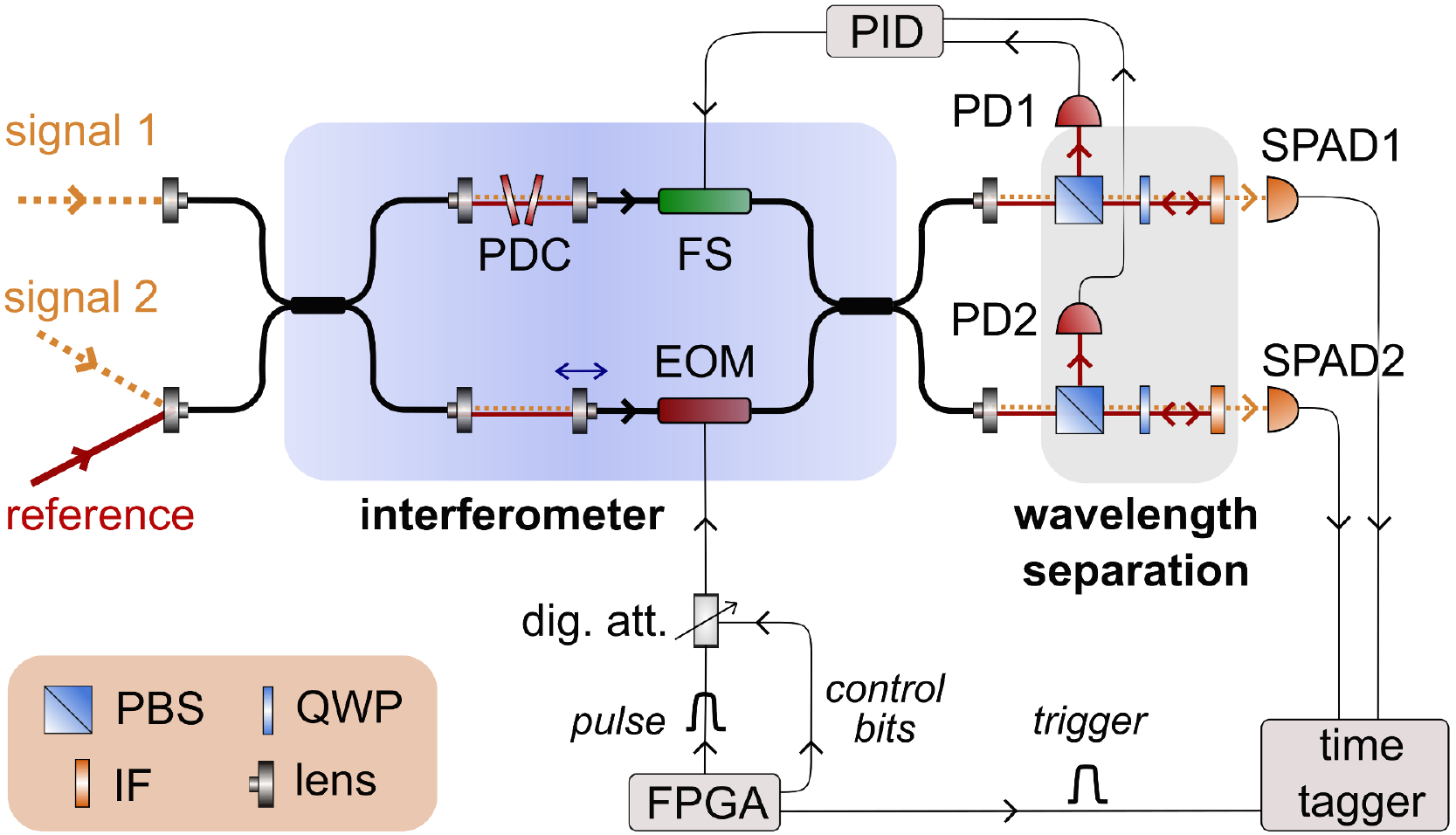}
\caption{Simplified experimental scheme of the switchable coupler showing the signal and reference beams entering the MZI and being separated at its output. Phase dispersion compensator (PDC) and fiber stretcher (FS) are used to lock the MZI phase based on the reference detection at the photodiodes (PD1,2). Integrated electro-optic modulator (EOM) driven by FPGA based electronics is employed for switching the coupler between arbitrary splitting ratios. The output signals are detected by single-photon avalanche diodes (SPAD1,2).}
\label{fig:experiment}
\end{figure}

The developed coupler is based on fiber MZI where the splitting ratio can be switched by changing an optical phase using an integrated waveguide EOM, see Fig.~\ref{fig:experiment}. The MZI was implemented to have high interference visibility resulting in high extinction ratio, exceptional phase stability enabling a long-term continuous operation, and fast modulation with low overall latency between a control electronic signal and the response of the switching. In what follows we will discuss these design goals and the corresponding features of the presented solution.

%% interference visibility %
High interference visibility operation requires perfect indistinguishability of interfering optical signals at the output of the MZI in all relevant degrees of freedom, namely path, spatial mode, polarization, time, and frequency. The path information is reduced by making the signals in both arms of the MZI of the same intensity by slight tuning of losses. Also, the splitting ratio of the output fiber splitter has to be close to 50:50. Spatial distinguishability is inherent in single-mode fiber implementation. Polarization-maintaining fibers are utilized throughout the setup to keep the polarization constant in time and the same for both the MZI arms. All connector splices are made to minimize polarization crosstalk between the fiber axes, and additional polarization filtering is also included. The MZI arms are carefully adjusted to have the same optical path length using tunable air gaps. The difference between the MZI arms is further minimized by placing the components symmetrically in both the arms. This is particularly important for the components exhibiting strong dispersion such as integrated EOMs and dispersion compensators. Having all these degrees of freedom under control and precisely adjusted, we have reached the interference visibility of 99.55\% in the optical bandwidth of 3~nm around 810~nm (equivalent to spectral bandwidth of 1.3 THz and pulse length down to 300 fs). It results in switching with the extinction ratio of 26~dB for continuous as well as pulsed optical signals.

%% phase stability %%
A usual problem of interferometric circuits is a random phase fluctuation caused by temperature changes, airflow, and vibrations. These adverse effects can be only partially reduced using passive methods such as thermal stabilization and acoustic and vibration isolation. On-chip circuit implementations exhibit improved stability, however, a few degree drift per minute is still present. Therefore an active stabilization is necessary to keep the phase fluctuation small enough for advanced applications. Particularly, long-term operation of the photonic routing circuit with the ultimate extinction ratio requires the phase stability better than 1~deg. Comparing the output signal to a fixed setpoint and adjusting the phase based on the error signal represents a common solution in the case of the strong classical signal. However, such an approach is fundamentally limited by a photocounting noise when a weak optical signal is used \cite{Huntington2005}. Inherently stable interferometers \cite{Micuda2014} or repeating the stabilization and measurements steps \cite{Mikova2012} are possible solutions at the single-photon level. The best performing technique uses an auxiliary strong optical reference co-propagating with the signal through the MZI and enabling the real-time phase lock \cite{Zeilinger2011}. In fibers, the reference and the signal overlap spatially and have to be multiplexed in different degree of freedom with the wavelength being the typical choice \cite{Xavier2015}.
We utilize 100~pW reference at 830~nm obtained from a spectrally and single-mode filtered luminescent diode. Its large spectral width of 10~nm allows for locking not only the optical phase but also the autocorrelation maximum, which signifies zero relative optical path of the interferometer. The reference is separated at the output of the MZI using a sequence of a polarizing beam splitter (PBS), quarter wave plate (QWP), and a 3~nm interference filter centered at 810~nm (IF) acting together as wavelength selective optical isolator. The transmitted signal is detected by single-photon avalanche diodes while the reflected reference impinges an ultra-sensitive photodiode (PD) with NEP=9 fW/$\!\sqrt{\mbox{Hz}}$. The amplified photodiode signal from both the MZI output ports is processed by a custom proportional-integral-derivative (PID) controller with the setpoint set at the maximum fringe slope and adaptively corrected for amplitude fluctuations of the reference. The produced electronic error signal is fed to a fiber stretcher (FS) with the dynamic range of 35~$\mu$m.
%The overall stabilization bandwidth of 1~kHz is given primarily by the response of the stretcher. During the measurements it was decreased to 30 Hz to reduce high-frequency noise of detectors.
The stabilization bandwidth was set to 30~Hz. The crosstalk from the reference to the signal is below 10 photons/s (i.e. photon crosstalk probability below $10^{-6}$ for 100~ns time bin). To manipulate the relative phase between the reference (locked to $\pi/2$) and the signal, we insert in the MZI a custom-made dispersion compensator formed by two tilted high-dispersion SF10 glass plates.

%% low latency %%
The response time, also termed latency, of the realized switchable coupler is given by the propagation delay of the optical signal from the input to the output of the device and, also, by the response of the phase modulator employed. The coupler is approximately 9~m long, which corresponds to the delay of 45~ns. It can be decreased below 10~ns easily by reducing the pigtail length of the constituent components and shortening the fiber stretcher sacrificing its dynamic range. Further decreasing the latency of the device seems to be superfluous especially when triggered by free-running single photon detectors considering their typical recovery time 10-30~ns. The waveguide integrated LiNbO${}_3$ EOM features 10~GHz bandwidth with negligible impact on the overall latency. The modulator is controlled by voltage signals within 0 -- 2.2~V using electronic pulse generator with 3.5~ns pulse width and 0.4~ns rise time for the response characterization, and a field-programmable gate array (FPGA) with 10~ns clock period to control complex measurement protocols. The FPGA was supplemented with a 6-bit digital attenuator with the 0.5~dB step to generate pulse sequences used for switching the coupler between arbitrary splitting ratios.

We have verified the stability of the splitting ratio during continuous wave operation and characterized the time response of the coupler to a fast-changing control signal, to demonstrate the outstanding performance of the developed coupler. The long-term stability was characterized by acquiring the output intensity for various fixed splitting ratios, particularly the most sensitive 50:50 ratio. Noise spectrum of the coupler transmittance shows 60~dB improvement for the actively real-time stabilized coupler. Allan deviation reaches the value of $5\times 10^{-4}$ for sub-second acquisitions times (affected by detector fluctuations) and exhibits a plateau at $10^{-4}$ for longer measurement durations. This is equivalent to phase stability of 0.6~deg, i.e. the optical path difference of the coupler's core interferometer is kept smaller than 1.5 nm for dozens of minutes.

\begin{figure}[t]
\centering
\includegraphics[width=0.95\linewidth]{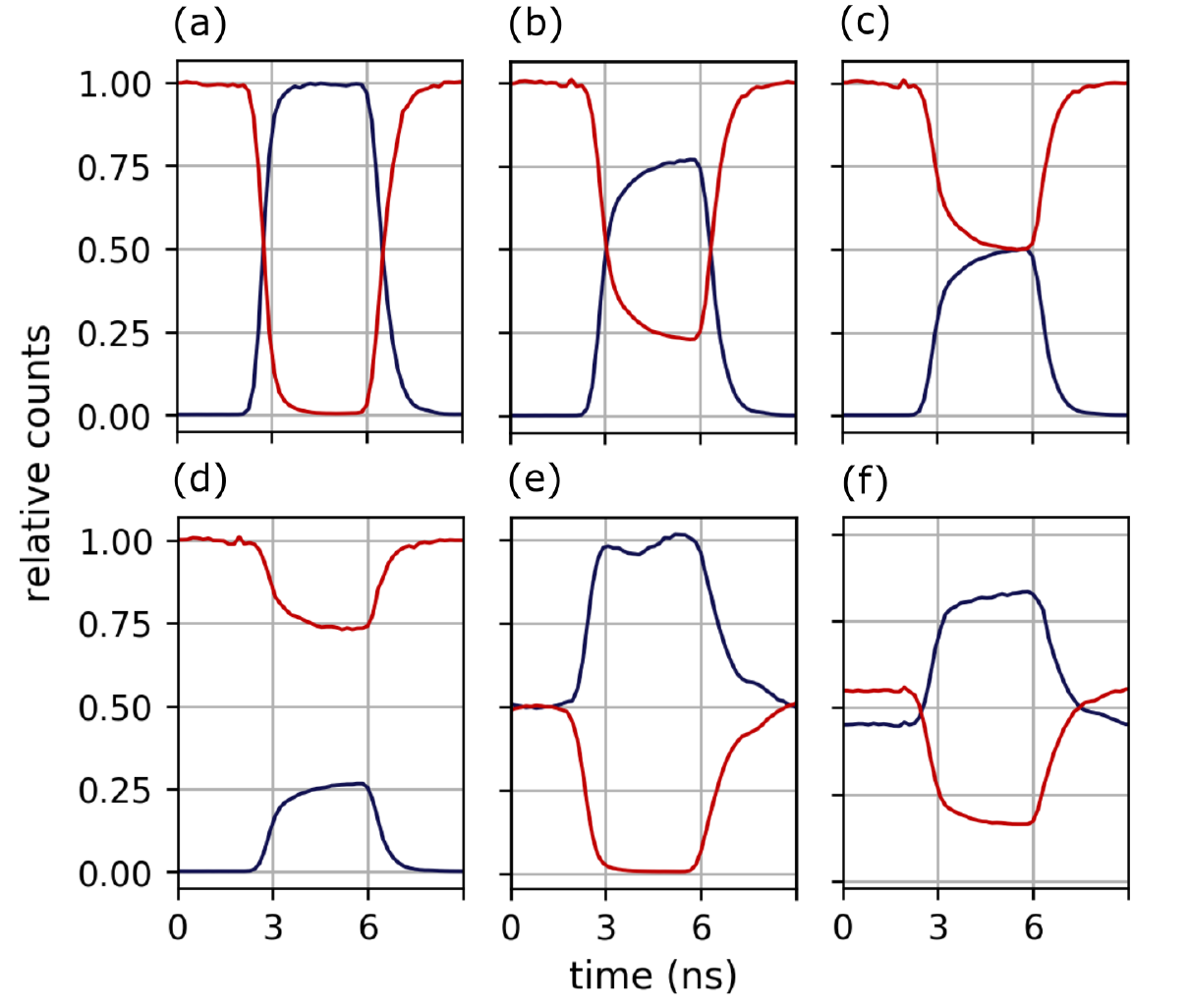}
\caption{The examples of fast switching with the splitting ratios: 100:0$\rightarrow$0:100 (a), 100:0$\rightarrow$25:75 (b), 100:0$\rightarrow$50:50 (c), 100:0$\rightarrow$75:25 (d), 50:50$\rightarrow$0:100 (e), and 55:45$\rightarrow$17:83 (f). Red and blue data points correspond to two outputs of the interferometer. The error bars are smaller than the data points.}
\label{fig:histograms}
\end{figure}

The time response was evaluated by setting a fixed initial splitting ratio and sending an electronic control pulse to the coupler. The switching process was observed at the output ports while the single-photon level signal was injected in the first input port of the device. The measurement was repeated many times due to the random nature of the photon detection process, and all detection events were recorded on a time-tagger. The accumulated photon-counting histograms are shown in Fig.~\ref{fig:histograms} for various initial and target splitting ratios to demonstrate arbitrariness of the switching. The data are depicted without corrections, except for single-photon avalanche diode (SPAD) afterpulses subtraction (maximum 1\% of the signal) and normalization, to show the temporal evolution of the transmittance and reflectance.
%The slight variability of the pulse shape is caused by the electronic pulse generator and RF components such as splitters and attenuators used for producing the control pulse with a proper amplitude.
The switching speed determined as the rise time (10\%-90\%) of the measured histograms is 0.7~ns, though the actual response of the coupler switching is much faster. The measurement is affected by the SPAD jitter (0.3 ns), the rise time of the electronic control pulse (0.4 ns), and a resolution of the time tagger (0.16 ns). After correcting for these contributions, the rise time of the coupler switching is estimated to be less than 100 ps, which is compatible with the integrated EOM speed of 10~GHz.

\begin{figure}[t]
\centering
\includegraphics[width=0.95\linewidth]{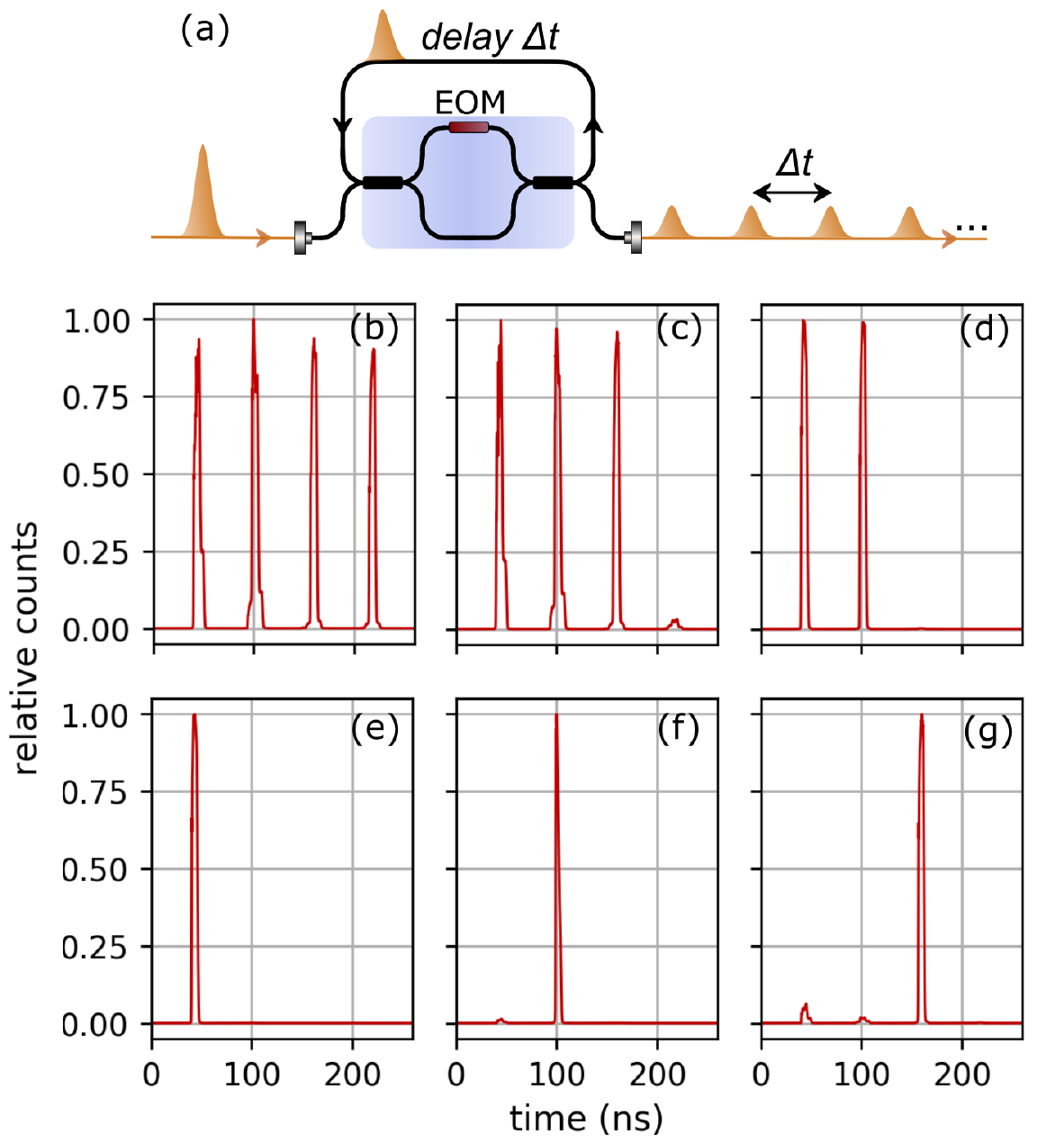}
\caption{Loop-based temporal multiplexing using the reported switchable coupler: scheme of the experiment (a) and various configurations of time-bin encoded 4-level photonic signal (b-g). Particularly, the panel (b) shows the balanced operation suitable for time-multiplexed photon-number-resolving detection.}
\label{fig:pulses}
\end{figure}

The input optical pulse can be multiplexed in many time bins when reflected part of the signal is fed from the output of the coupler to its input to create a loop, as shown in Fig.~\ref{fig:pulses}(a). Electronic control pulses applied to the EOM have to be synchronized with the optical pulse repeatedly passing the coupler. This scheme follows the proposal of a time-multiplexed device for photon-number-resolving detectors \cite{Banaszek2003}.
Recently, the scheme was experimentally verified, employing a binary switch based on a free-space Pockels cell with the latency of 2.4~$\mu$s corresponding to a fiber delay loop length of 480~m \cite{Silberhorn2019}.
The utilized fixed splitting ratio switching results in non-uniform probability distribution of finding a photon in individual time bins.
Employing the tunable coupler reported here, we were able to reach the fully balanced operation and, at the same time, decrease the latency to 60~ns corresponding to a 12~m long fiber, i.e. a direct connection between the output and input pigtails of the coupler.

The reported switching protocol can be generalized to arbitrary time multiplexing. We demonstrate full control over the amplitude of the individual time bins with the mean fidelity of 98.9\%. Several examples of time-bin encoded 4-level optical system are depicted in Fig.~\ref{fig:pulses}(b-g). The tunable routing of the input signal to the resulting time bins can be complemented by their arbitrary phase modulation using EOMs in both the MZI arms. Starting from single photon input, such the routing represents an efficient way of preparing a photonic multi-level system (qudit). A second switchable coupler would be needed for the qudit analysis at the receiver.

The overall loss of the coupler and the loop represents the main limitation of a photon-number-resolving loop-detector, as the signal is diminished in each cycle in the loop. The extinction ratio of the coupler determines the minimum probability of releasing a photon before the first full cycle. Here, we have focused on the extinction ratio and latency of the coupler and not performed an extensive loss optimization; hence the total loss during a single cycle is approximately 80\%. It limits the multiplexing to four balanced time bins, see Fig.~\ref{fig:pulses}(b). Using the same fiber architecture with low loss off-the-shelf components, the total loss could be decreased down to approximately 50\% with the main contribution stemming from the integrated EOM, which corresponds to 8 balanced round trips. Further loss reduction is expected with the promising development of integrated phase modulators. Thin-film lithium-niobate EOM with 100 GHz bandwidth and on-chip loss of 0.1 dB was reported recently \cite{wang2018integrated}, though further reduction of the fiber-to-chip coupling loss and improvement of the optical bandwidth would be necessary to reach the presented coupler performance.
%The most efficient coupling exhibits 0.58 dB loss per facet, and yet was achieved only on SOI platform that exhibits relatively high on-chip loss \cite{ding2014fully,marris2013low}.
On-chip implementation of optical delay loops required for loop-based detectors and computing represents
another significant challenge. Alternatively, a free-space configuration of the whole circuit can be adopted using bulk modulators and delay lines with estimated overall loss slightly below 10\%. We estimate time multiplexing to 30-40 of non-saturated {\em equiprobable} channels to be ultimately possible. It might be challenging, however, to reach very low latency and large electronic bandwidth due to high driving voltage required by bulk electro-optic modulators.
Ultimately, the response time can be reduced below 150 ns for discrete switching \cite{Pan2017full,takeda2019demand}.

It is important to stress that the inevitable presence of losses does not prevent utilization of the switchable coupler in a range of single-photon application in the field of quantum technology. Except few rare measurement protocols, e.g. conditioning on the vacuum state, the finite efficiency can be corrected for. The quality of the produced quantum state or quantum transformation is not typically decreased by losses but the rate of the process is affected. Particularly, in applications employing measurement in coincidence basis \cite{Sciarrino2019,Pryde2019} the reported coupler will perform nicely despite its nonzero loss. 

We have presented the sub-nanosecond switchable coupler optimized for routing faint optical signals and single photons. The measured overall latency of the coupler is 45 ns with a possibility of reduction below 20 ns, which is comparable with the recovery time of the state-of-the-art single-photon detectors. We have verified full tunability of the splitting ratio from 0:100 to 100:0 with the exceptional extinction of 26 dB and unparalleled long-term stability of one part in 10,000.
We have reached for the first time the balanced operation of loop-based photon-number-resolving detector exploiting the full control over the splitting ratio and ultimate stability of the developed coupler. Furthermore, we have demonstrated the deterministic preparation of photonic time-bin four-level qudit with a clock cycle of 60 ns using the presented coupler and a single delay loop.
We envision the use of the reported device in advanced feedback and feedforward based schemes of electro-optical control of light, where a detection of a fraction of the light signal changes the splitting ratio of the remaining signal. The low-latency switchable coupler is the key device instantly applicable in a vast number of applications such as time-multiplexed single-photon sources \cite{OBrien2016}, photon-number-resolving detectors \cite{Silberhorn2019}, and various time-bin encoded communication protocols \cite{Vallone2018} including quantum key distribution \cite{Gauthier2018} and hyper-entangled states preparation and measurement \cite{Predojevic2018}.

\section*{Funding Information}
Czech Science Foundation (project 17-26143S);
MEYS and European Union's Horizon 2020 (2014-2020) research and innovation framework programme under grant agreement No 731473 (QuantERA project HYPER-U-P-S No 8C18002);
Palack\'y University (project IGA-PrF-2019-010).

\section*{Disclosures}
The authors declare no conflicts of interest.
%\section*{Acknowledgments}
%This work was supported by the Czech Science Foundation (project 17-26143S). MJ acknowledges the support of QuantERA project HYPER-U-P-S funded from the MEYS and European Union's Horizon 2020 (2014-2020) research and innovation framework programme under grant agreement No 731473 (project No 8C18002). VŠ acknowledges the support by the Palack\'y University (project IGA-PrF-2019-010).

%\section*{Supplemental Documents}
%\bigskip \noindent See \href{svarcsupplement}{Supplement 1} for supporting content.
%\emph{Optica} authors may include supplemental documents with the primary manuscript. For details, see \href{http://www.opticsinfobase.org/submit/style/supplementary-materials-optica.cfm}{Supplementary Materials in Optica}. To reference the supplementary document, the statement ``See Supplement 1 for supporting content.'' should appear at the bottom of the manuscript (above the references).

%% Bibliography
%\bibliography{bibliography_switch}
%% 5. strana s plnýma referencema je potřeba, měla by se generovat automaticky, když se používá bibtex - opt. letters to prý umožňuje

%

\end{document}